\documentclass[reprint,preprintnumbers,superscriptaddress,amsmath,amssymb,showkeys]{revtex4-1}
\usepackage{hyperref}
\usepackage{float}
\usepackage{natbib}
\usepackage{graphicx}
\usepackage{dcolumn}
\usepackage{bm}
\usepackage{siunitx} 

\begin{document}

\bibliographystyle{unsrtnat}
\title{\large {Observation of Intervalley Biexcitonic Optical Stark Effect in Monolayer WS$_2$}}

\vspace{0.5cm}
\author{Edbert J. Sie}
		\affiliation{Department of Physics, Massachusetts Institute of Technology, Cambridge, MA 02139, USA}
\author{Chun Hung Lui}
		\affiliation{Department of Physics and Astronomy, University of California, Riverside, California 92521, USA}
\author{Yi-Hsien Lee}
		\affiliation{Materials Science and Engineering, National Tsing-Hua University, Hsinchu 30013, Taiwan}
\author{Jing Kong}
		\affiliation{Department of Electrical Engineering and Computer Science, Massachusetts Institute of Technology, Cambridge, MA 02139, USA}		
\author{Nuh Gedik}
		\altaffiliation{gedik@mit.edu}
		\affiliation{Department of Physics, Massachusetts Institute of Technology, Cambridge, MA 02139, USA}
\vspace{0.5cm}

\begin{abstract}
Coherent optical dressing of quantum materials offers technological advantages to control their electronic properties, such as the electronic valley degree of freedom in monolayer transition metal dichalcogenides (TMDs). Here, we observe a new type of optical Stark effect in monolayer WS$_2$, one that is mediated by \textit{intervalley biexcitons} under the blue-detuned driving with circularly polarized light. We found that such helical optical driving not only induces an exciton energy downshift at the excitation valley, but also causes an anomalous energy upshift at the opposite valley, which is normally forbidden by the exciton selection rules but now made accessible through the intervalley biexcitons. These findings reveal the critical, but hitherto neglected, role of biexcitons to couple the two seemingly independent valleys, and to enhance the optical control in valleytronics.
\end{abstract}


\maketitle

Monolayer TMDs host tightly-bound excitons in two degenerate but inequivalent valleys (K and K$^{\prime}$), which can be selectively photoexcited using left ($\sigma^-$) or right ($\sigma^+$) circularly polarized light (Fig 1a) \cite{Xiao12, Mak12, Zeng12, Cao12}. The energy levels of these excitons can be tuned optically in a valley-selective manner by means of the optical Stark effect \cite{Sie15, Kim14}. Prior research has demonstrated that monolayer TMDs driven by below-resonance (red-detuned) circularly polarized light can exhibit an upshifted exciton level, either at the K or K$^{\prime}$ valleys depending on the helicity, while keeping the opposite valley unchanged. This valley-specific phenomenon arises from the exciton state repulsion by the photon-dressed state in the same valley, a mechanism consistent with other optical Stark effects in solids \cite{Mysyrowicz86, VonLehmen86}.

\begin{figure}[t]
	\includegraphics[width=0.48\textwidth]{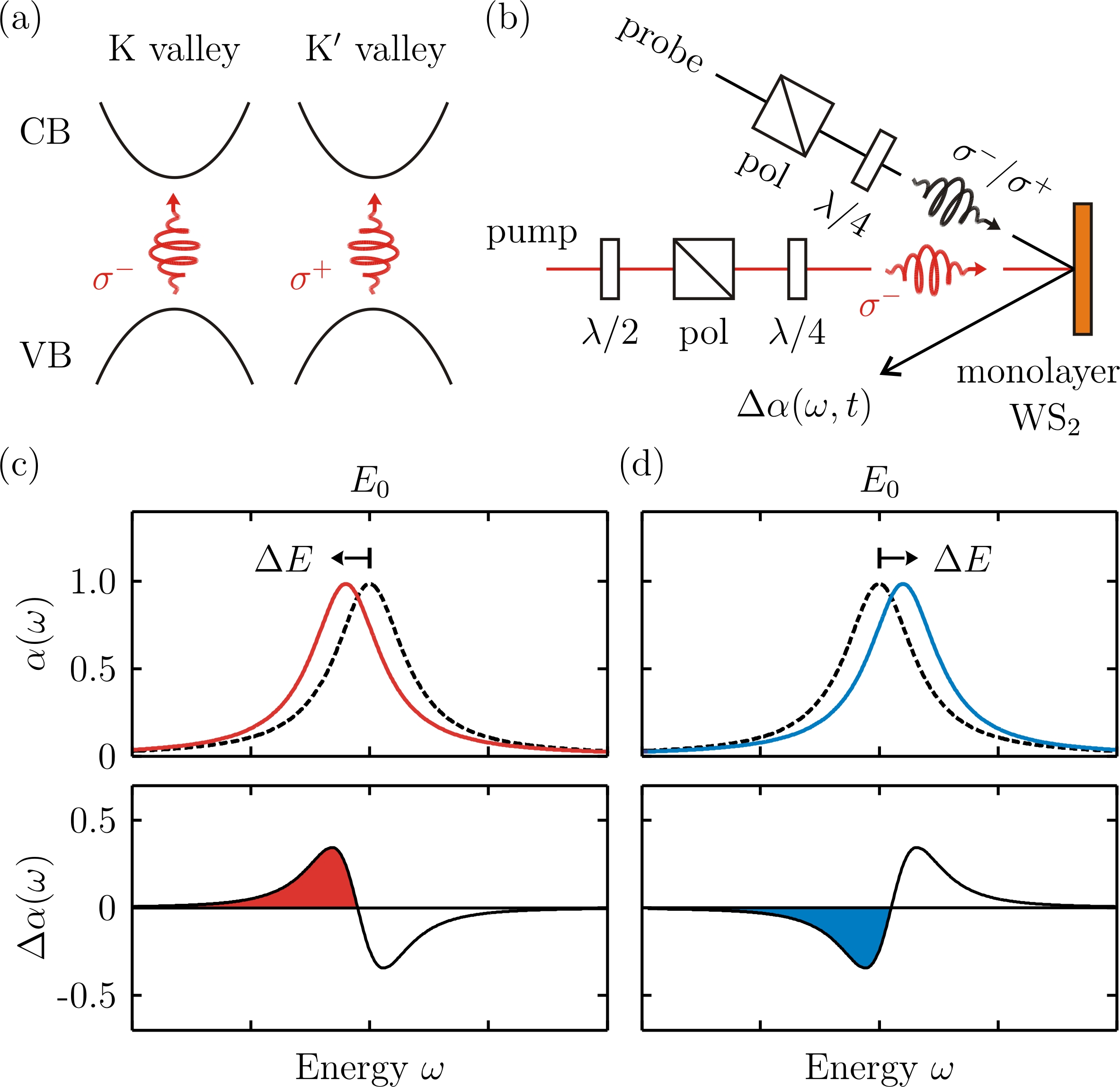}
	\caption{(a) K and K$^{\prime}$ valleys couple selectively with left ($\sigma^-$) and right ($\sigma^+$) circularly polarized light due to selection rules. (b) Schematic of the pump-probe spectroscopy setup. (c-d) Simulated  absorption spectra $\alpha (\omega)$ that are shifted by $\Delta E$ to lower and higher energies (upper panels), as well as their induced absorption spectra $\Delta\alpha (\omega)$ (lower panels).}
	\label{fig:Fig1}
\end{figure}

Despite much recent progress, a complete understanding of the optical Stark effect in monolayer TMDs is still lacking. First, the anticipated complementary effect of using above-resonance (blue-detuned) light to downshift the exciton level has not been demonstrated. This is challenging because the blue-detuned light excites real exciton population, which can easily obscure the optical Stark effect. Secondly, when the detuning is sufficiently small and comparable to the biexciton binding energy, the effect may involve a coherent formation of the recently identified intervalley biexcitons \cite{Mai14, Sie15B}. These biexcitons are expected to have profound contributions to the optical Stark effect, as indicated by earlier studies in semiconductor quantum wells \cite{Combescot88, Hulin90}. Elucidating these processes is therefore crucial to investigate the role of intervalley biexcitons in monolayer TMDs in order to obtain a thorough understanding of the coherent light-matter interactions in this system.

\begin{figure*}[t]
	\includegraphics[width=0.85\textwidth]{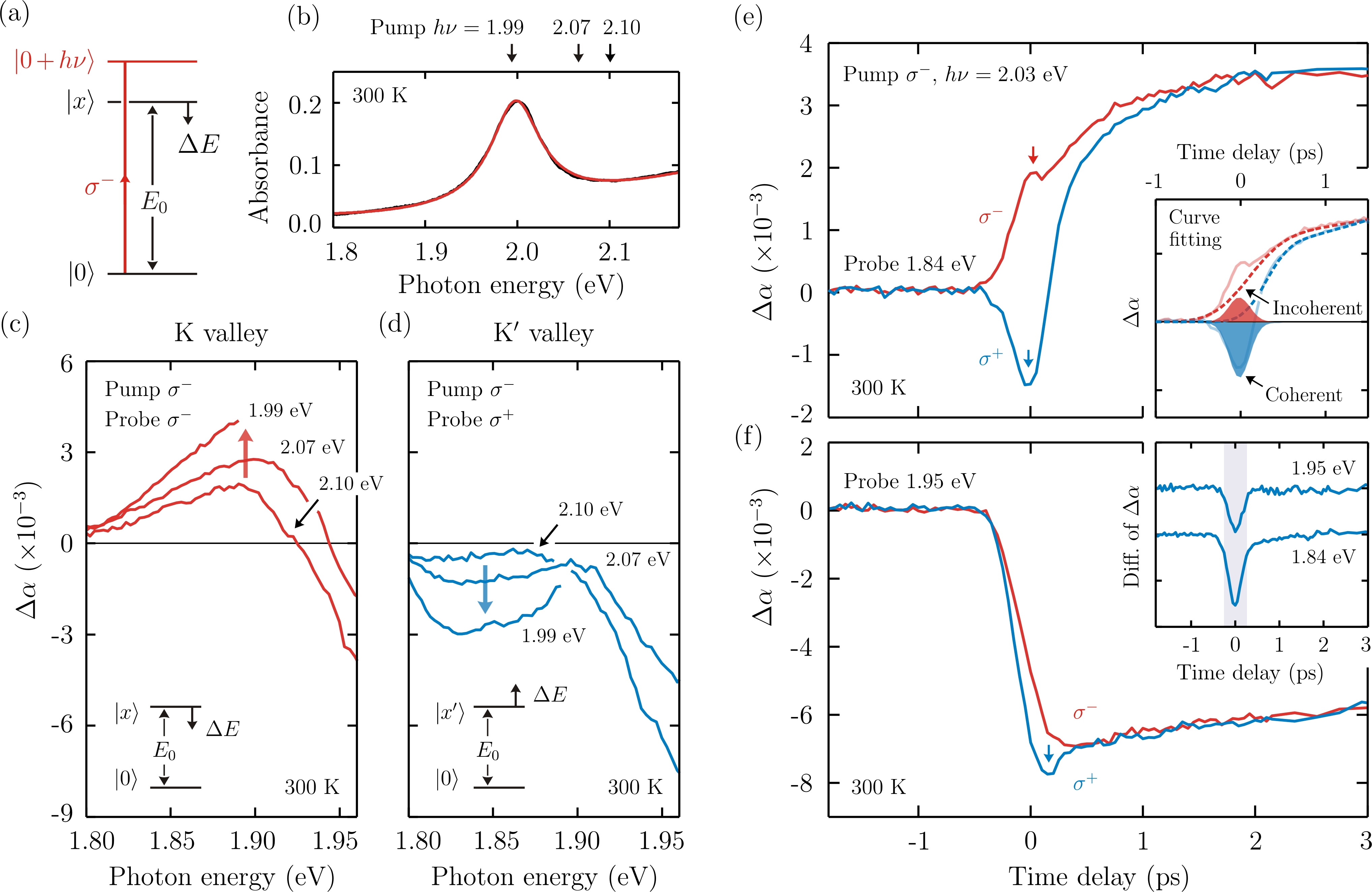}
	\caption{Blue-detuned optical Stark effect and its observation in monolayer WS$_2$. (a) Blue-detuned optical driving scheme, where we use $\sigma^-$ pump pulse with photon energy $h\nu$ slightly above the exciton resonance $E_0$. (b) Measured absorption spectrum of monolayer WS$_2$ shows that $E_0 = 2.00$ eV at 300 K. (c-d) Valley-specific $\Delta\alpha$ spectra induced by $\sigma^-$ pump pulses ($h\nu = 1.99, 2.07, 2.10$ eV) and monitored by using $\sigma^-$ (K) and $\sigma^+$ (K$^{\prime}$) broadband probe pulses at pump-probe time delay $\Delta t = 0$. The increasing $\Delta\alpha$ at K valley indicates a pump-induced redshift of exciton energy. On the other hand, the decreasing $\Delta\alpha$ at K$^{\prime}$ valley, though unexpected, indicates a pump-induced blueshift of exciton energy. (e-f) Time traces of $\Delta\alpha$ induced by $\sigma^-$ pump pulses ($h\nu = 2.03$ eV) and monitored at probe energy of 1.84 eV and 1.95 eV with different helicities. The top inset shows the curve fitting decomposition of the coherent and incoherent signals. The bottom inset shows the valley contrast of the signals, $\Delta\alpha (\sigma^+) - \Delta\alpha (\sigma^-)$, where the two curves are offset for clarity.}
	\label{fig:Fig2}
\end{figure*}

In this letter, we explore the optical Stark effect using blue-detuned optical driving in monolayer TMD WS$_2$. We found that by driving the system using intense laser pulses with blue-detuned and left circular polarization, we can \textit{lower} the exciton energy at the K valley. In addition, as the driving photon energy approaches the resonance, an unexpected and remarkable phenomenon emerges -- the exciton energy at the opposite (K$^{\prime}$) valley is \textit{raised}. This observation is anomalous because interaction with the driving photon is forbidden at this valley by the exciton selection rules. The upshifted exciton level also contrasts sharply with the downshifted level at the K valley. These findings reveal the strong influence of intervalley biexcitons to the optical Stark effect. By including their contributions in an expanded four-level optical Stark effect, we are able to account for all the main observations in our experiment.

We monitor the pump-induced change of exciton levels at the K (K$^{\prime}$) valley using the reflection of synchronized broadband probe pulses with $\sigma^-$ ($\sigma^+$) polarization (Fig 1b, see Supporting Information). The sample consists of high-quality monolayers of WS$_2$ that were grown by chemical vapor deposition on sapphire substrates \cite{Lee12, Lee13, Gutierrez12}. For such an atomically-thin layer on a transparent substrate, the change of absorption ($\Delta\alpha$) can be directly extracted from the change of reflection \cite{Sie15B}. The resulting $\Delta\alpha$ spectrum allows us to determine the direction and magnitude of the exciton energy shift ($\Delta E$), which typically shows a single-cycle waveform, as depicted in Fig 1c-d. In our experiment, we examine the lower-energy part of $\Delta\alpha$ spectrum (filled color in Fig 1c-d), because the coherent contribution is more pronounced below the energy resonance ($E_0$). For a blue-detuned optical Stark effect (Fig 2a), the pump photon energy is tuned to be slightly higher than the exciton resonance in monolayer WS$_2$, which is at $E_0 = 2.00$ eV from our measured absorption spectrum (Fig 2b) as well as from other experiments \cite{LiPRB14}.

Figure 2c shows the $\Delta\alpha$ spectra at zero pump-probe delay at three different pump photon energies ($h\nu =$ 2.10, 2.07, 1.99 eV) but the same pump fluence (\SI{28}{\micro\J}/cm$^2$). We display the spectra in the range of 1.80$-$1.96 eV, where the coherent effect is more pronounced and less contaminated by the pump scattering. For the $\sigma^-$ probe (Fig 2c), the spectral shape is similar to that in Fig 1c, indicating a redshift of the exciton level at the same (K) valley. As the pump photon energy approaches the resonance from 2.10 to 1.99 eV, the magnitude increases considerably, indicating an increasing redshift of the exciton level. This observation complements the previous studies, which reported a blueshift using red-detuned pump pulse. In contrast, the $\Delta\alpha$ spectra at the opposite (K$^{\prime}$) valley exhibit a distinct form, as revealed by the $\sigma^+$ probe (Fig 2d). Its value is negative in the range of 1.80$-$1.90 eV, with a waveform similar to that in Fig 1d. This indicates a blueshift of the exciton level at the K$^{\prime}$ valley, which becomes more substantial as the pump approaches the resonance.

\begin{figure}[t]
	\includegraphics[width=0.40\textwidth]{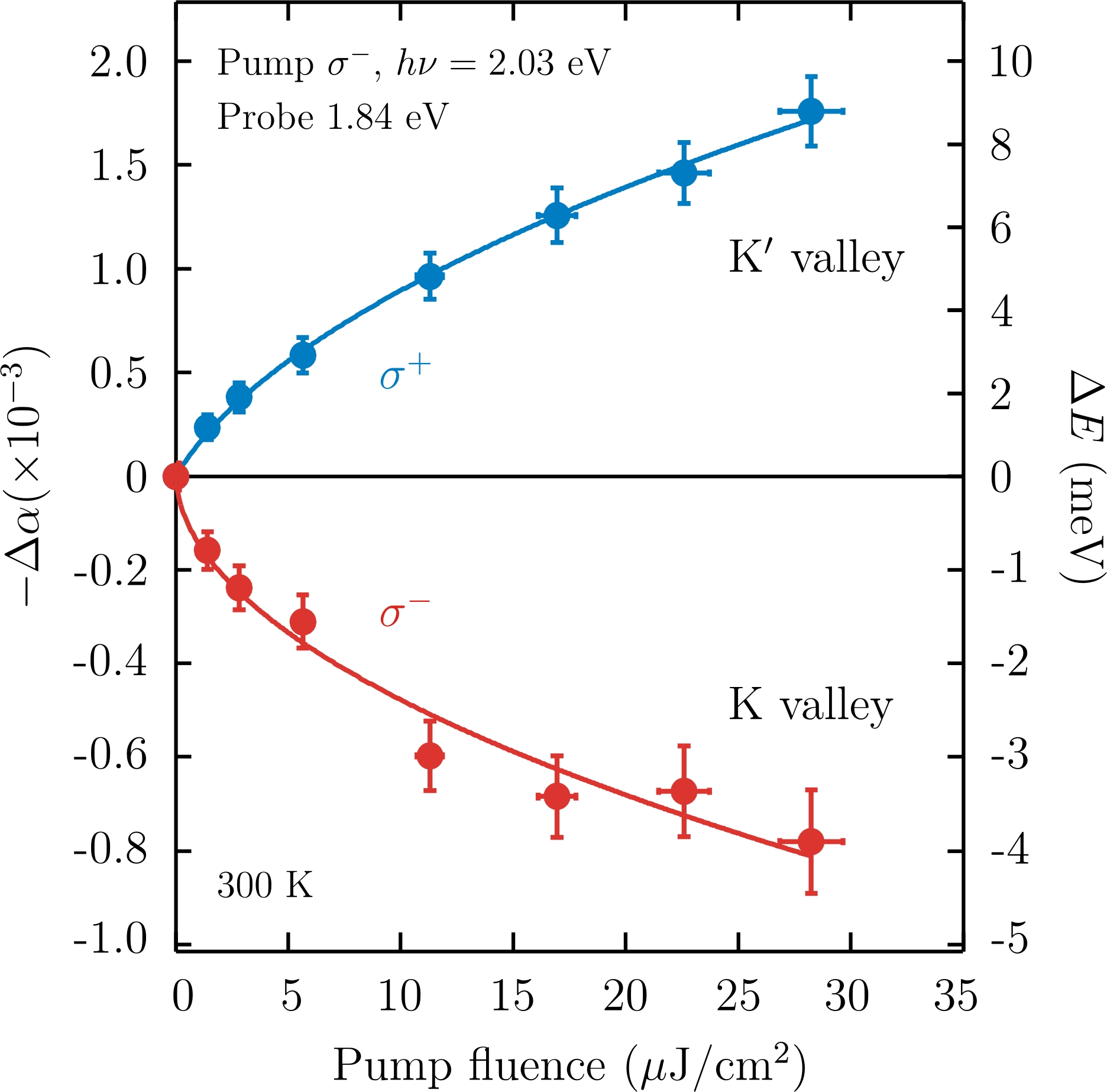}
	\caption{Fluence dependence of the blue-detuned optical Stark shift. The measured data of $-\Delta\alpha (\propto \Delta E)$ are plotted at increasing pump fluence ($\sigma^-$, $h\nu = 2.03$ eV), measured valley-selectively at probe energy of 1.84 eV. The energy scale on the right axis is estimated based on the measured absorption slope of 0.2/eV at 1.84 eV. Note that the energy scaling is different between the two valleys. The fitting curves show that the K and K$^{\prime}$ valleys exhibit square-root fluence dependences, as discussed in the main text.}
	\label{fig:Fig3}
\end{figure}

We note that the observed spectra include two types of contributions -- the coherent and incoherent signals. The coherent signal arises from the optical Stark effect. The incoherent signal arises from the created exciton population, which is unavoidable using the above-resonance photoexcitation, causing band renormalization, biexciton absorption and Pauli blocking \cite{Mai14, Sie15B, Mai14B, Wang15, Schaibley15, Wang13, Sim13, Shi13, Pogna16, Steinhoff14}. These two types of processes evolve differently with the pump-probe time delay. The coherent signal appears only within the pump pulse duration, whereas the incoherent signal remains after the pulsed excitation. Using such distinct time dependence, we have separated the coherent signal from the incoherent background by monitoring the $\Delta\alpha$ time traces. Figure 2e-f shows the time traces, induced by pump pulses with energy $h\nu = 2.03$ eV and duration 200 fs. At finite pump-probe delay ($\Delta t > 1$ ps), $\Delta\alpha$ is similar for both valleys, with positive value at around 1.84 eV (Fig 2e) but negative value at around 1.95 eV (Fig 2f). These features correspond to the exciton population effects. At zero pump-probe delay, however, the two valleys exhibit significantly different response. The difference can be attributed to the optical Stark effect, a coherent process that follows the pump pulse intensity profile. At probe energy 1.84 eV, the coherent contribution is particularly prominent and can be readily separated from the incoherent background by direct extrapolation (Insets of Fig 2e-f and Supporting Information).

We have extracted the coherent component of $-\Delta\alpha$ at 1.84 eV and plot the values as a function of pump fluence (Fig 3). The associated energy shift $\Delta E$ can be estimated from the differential form $\Delta\alpha(\omega,\Delta E) = - (d\alpha/d\omega)\Delta E$ \cite{derivation}. Given the measured $-\Delta\alpha$ and the slope at 1.84 eV, we have evaluated such energy shift (the right vertical axis of Fig 3). Our result shows that the exciton level at K and K$^{\prime}$ valleys respectively downshifts ($-4$ meV) and upshifts ($+9$ meV) under the $\sigma^-$ blue-detuned optical driving. The magnitude of both shifts increases sublinearly with pump fluence, in contrast to the linear fluence dependence in prior red-detuned experiments.

The upshift of the exciton level at K$^{\prime}$ valley is anomalous. First, according to the well-known selection rules in this system, the K$^{\prime}$ valley is not accessible by the $\sigma^-$ (K valley) optical driving. The observed optical Stark effect at K$^{\prime}$ valley apparently violates this selection rule. Secondly, even if the access to the K$^{\prime}$ valley is allowed, a blue-detuned optical driving is expected to downshift the exciton level, as in the case of the K valley. The energy upshift at the K$^{\prime}$ valley apparently defies this common knowledge of optical Stark effect, hence it must arise from a different mechanism, one that is beyond the framework of interaction between light and single excitons.

\begin{figure}[t]
	\includegraphics[width=0.40\textwidth]{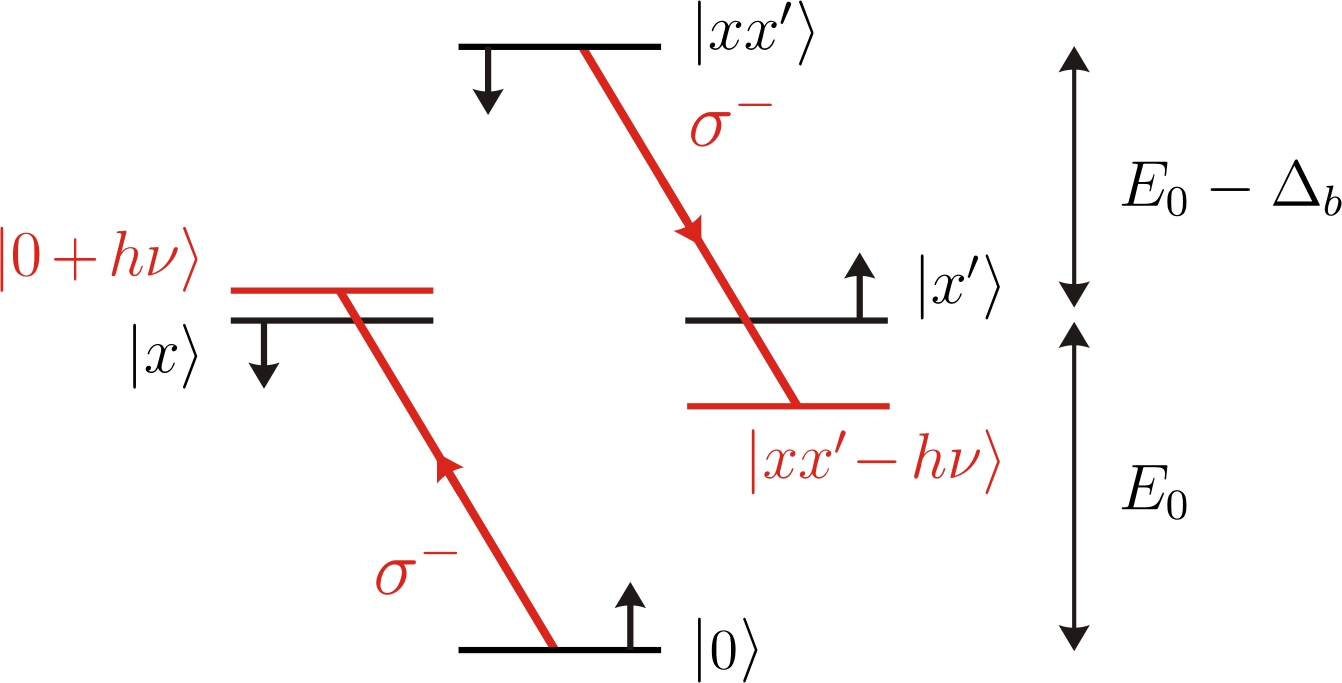}
	\caption{Energy level diagram of the intervalley biexcitonic optical Stark effect. Here the $\sigma^-$ pump pulse is blue-detuned, above the energy resonance between the ground state $\left|0\right\rangle$ and the exciton state $\left|x\right\rangle$. Coherent absorption from $\left|0\right\rangle$ results in a photon-dressed state $\left|0+h\nu\right\rangle$, while coherent emission from the intervalley biexciton state $\left|xx^{\prime}\right\rangle$ results in a photon-dressed state $\left|xx^{\prime}-h\nu\right\rangle$.}
	\label{fig:Fig4}
\end{figure}

We interpret this observation as resulting from the optical Stark effect that is mediated by intervalley biexcitons. Recent research has revealed significant interactions between individual excitons in monolayer TMDs. In particular, two excitons at different valleys can be bound to form an excitonic molecule, the intervalley biexciton, with unusually large binding energies (40$-$70 meV) \cite{Mai14, Sie15B, Shang15, You15, Kylanpaa15}. These intervalley biexcitons offer an effective channel to couple the two valleys, with selection rules different from those for single excitons. In view of such strong biexcitonic effect, we can account for our observations within a four-level scheme, which includes the ground state $\left|0\right\rangle$, the two valley exciton states $\left|x\right\rangle$ and $\left|x^{\prime}\right\rangle$, and the intervalley biexciton state $\left|xx^{\prime}\right\rangle$ (Fig 4). In this scheme, the optical pumping creates two types of photon-dressed states -- one from the ground state $\left|0+h\nu\right\rangle$ and the other from the biexciton state $\left|xx^{\prime}-h\nu\right\rangle$. The former can interact with the exciton state $\left|x\right\rangle$ at the K valley. Since $\left|0+h\nu\right\rangle$ lies above $\left|x\right\rangle$ in a blue-detuned experiment, repulsion between the two states causes $\left|x\right\rangle$ to downshift. This is responsible for the normal optical Stark effect at the K valley (red dots in Fig 3). In contrast, the biexciton photon-dressed state $\left|xx^{\prime}\right\rangle$ can interact with the exciton state $\left|x^{\prime}\right\rangle$ at the opposite (K$^{\prime}$) valley according to different selection rules for the intervalley biexciton. Since $\left|xx^{\prime}-h\nu\right\rangle$ lies below $\left|x^{\prime}\right\rangle$, repulsion between the two will cause $\left|x^{\prime}\right\rangle$ to upshift. This is responsible for the anomalous optical Stark effect at the K$^{\prime}$ valley (blue dots in Fig 3). It is evident that the intervalley biexciton plays a unique role in coupling the two valleys, and the effect can be utilized for enhanced control of valley degree of freedom \cite{Xu14}.

In order to investigate the photon-induced coupling between these states, we consider a four-level Jaynes-Cummings model, with a procedure similar to but extended from our previous work \cite{Sie15, Jaynes63}. Such a model has been successfully applied to describe the light-dressed states in many semiconductor systems, and can readily be adopted to describe the exciton-biexciton system \cite{Sie15, Mysyrowicz86, VonLehmen86, Koster11, delValle10, Schumacher12}. By virtue of the unique valley selection rules in this system, the originally 4$\times$4 Hamiltonian matrix can be simplified into two \textit{decoupled} 2$\times$2 Hamiltonian matrices
\begin{eqnarray}
\begin{split}
\hat{H}_K &= \frac{1}{2}E_0\hat{\sigma}_z + h\nu\hat{a}^{\dagger}\hat{a} + \frac{1}{2}g\left(\hat{\sigma}^{\dagger}\hat{a} + \hat{\sigma}\hat{a}^{\dagger}\right) \\
\hat{H}_{K^{\prime}} &= \frac{1}{2}\left(E_0-\Delta_b\right)\hat{\sigma}_z + h\nu\hat{a}^{\dagger}\hat{a} + \frac{1}{2}g^{\prime}\left(\hat{\sigma}^{\dagger}\hat{a} + \hat{\sigma}\hat{a}^{\dagger}\right) \\
\end{split}
\end{eqnarray}
The three terms in each Hamiltonian correspond to the two-level system, the photon reservoir, and the exciton-photon interactions, respectively. Possible contribution from real exciton population is neglected in this model. Here, $g$ and $g^{\prime}$ are the exciton-photon coupling strengths, $\hat{\sigma}$'s are the Pauli matrices, $\hat{a}$ and $\hat{a}^{\dagger}$ are the photon ladder operators, and $\Delta_b$ is the biexciton binding energy. The Hamiltonian $\hat{H}_K$ couples states $\left|0,n+1\right\rangle$ and $\left|x,n\right\rangle$, while $\hat{H}_{K^{\prime}}$ couples states $\left|x^{\prime},n\right\rangle$ and $\left|xx^{\prime},n-1\right\rangle$, where $\left|n\right>$ is the number of photons. By using these states as the basis (see Supporting Information), we can express the Hamiltonian matrices
\begin{eqnarray}
\begin{split}
H_K &= \frac{1}{2}
\left(
\begin{array}{cc}
h\nu-E_0 & g\sqrt{n+1} \\ 
g\sqrt{n+1} & -\left(h\nu-E_0\right) \\
\end{array}
\right)\\
H_{K^{\prime}} &= \frac{1}{2}
\left(
\begin{array}{cc}
h\nu-E_0+\Delta_b & g^{\prime}\sqrt{n} \\ 
g^{\prime}\sqrt{n} & -\left(h\nu-E_0+\Delta_b\right) \\
\end{array}
\right)
\end{split}
\end{eqnarray}
This is in addition to the photon reservoir terms, $h\nu(n+1/2)$ and $h\nu(n-1/2)$, which only contribute to the energy offsets. By diagonalizing the above matrices, we can obtain the energy levels of the photon dressed states $E_K = \pm \frac{1}{2} \sqrt{(h\nu-E_0 )^2+g^2 (n+1)}$ and $E_{K^{\prime}} = \pm \frac{1}{2} \sqrt{(h\nu-E_0+\Delta_b)^2+g^{\prime 2} (n)}$, where $g\sqrt{n+1} = \mathcal{M}\mathcal{E}_0$ and $g^{\prime}\sqrt{n} = \mathcal{M}^{\prime}\mathcal{E}_0$ are the Rabi frequencies. Here $\mathcal{M}$ and $\mathcal{M}^{\prime}$ are the moments for $\left|0\right\rangle \rightarrow \left|x\right\rangle$ and $\left|x^{\prime}\right\rangle \rightarrow \left|xx^{\prime}\right\rangle$ transitions, respectively, and $\mathcal{E}_0$ is the electric field amplitude of the light.

From these expressions, we can finally obtain the optical Stark shifts of the exciton levels
\begin{eqnarray}
\begin{split}
\Delta E_K &= -\frac{1}{2}\left(\sqrt{(h\nu-E_0)^2+\mathcal{M}^2\mathcal{E}_0^2}-(h\nu-E_0)\right) \\
\Delta E_{K^{\prime}} &= \frac{1}{2}\left(\sqrt{(h\nu-E_0+\Delta_b)^2+\mathcal{M}^{\prime 2}\mathcal{E}_0^2} - (h\nu-E_0+\Delta_b)\right)
\end{split}
\end{eqnarray}
Despite much similarity, the two optical Stark effects are quantitatively different, because the transition moments are generally different and the biexciton photon-dressed state is offset by $\Delta_b$. In the large detuning limit $h\nu - E_0 \gg \mathcal{M}\mathcal{E}_0$, we retrieve the well-known expression $\Delta E_K = -\mathcal{M}^2\mathcal{E}_0^2/4(h\nu-E_0)$ with a linear fluence dependence, as observed in the previous red-detuned experiment. Conversely, in the small detuning limit $h\nu-E_0 \ll \mathcal{M}\mathcal{E}_0$, we obtain $\Delta E_K = -\frac{1}{2}\sqrt{\mathcal{M}^2\mathcal{E}_0^2}$ with a square-root fluence dependence. The observed sublinear fluence dependence in Fig 3 indicates that the small-detuning limit is reached for both valleys in our experiment.  Our fluence dependence data can be fitted with this model (Fig 3), with transition moments and effective detunings as adjustable parameters (Supporting Information). The good agreement between the experiment and the model strongly supports that this optical Stark effect is mediated by intervalley biexcitons. 

In summary, we have observed an exciton energy downshift at the excitation (K) valley, and an energy upshift at the opposite (K$^{\prime}$) valley, under the blue-detuned optical driving in monolayer WS$_2$. While the energy downshift arises from the single-exciton optical Stark effect, the anomalous energy upshift is attributed to the intervalley biexciton optical Stark effect because it exhibits three characteristics: (i) It emerges only within the pump pulse duration, (ii) it has a square-root dependence on the pump fluence, and (iii) it obeys the biexcitonic valley selection rule for opposite circularly polarized light, consistent with our model. Our results show that the intervalley biexciton is not only a rare and interesting quasiparticle by itself, but it also plays an active role to channel a coherent and valley-controllable light-matter interaction.

Finally, apart from slight quantitative difference the two types of optical Stark effects exhibit beautiful contrast and symmetry with the valley indices (K, K$^{\prime}$) and the direction of the energy shift (down and up shifts). The optical Stark effect at K valley arises from \textit{intravalley} exciton-exciton interaction through statistical Pauli repulsion, whereas the effect at K$^{\prime}$ valley arises from \textit{intervalley} exciton-exciton interaction through biexcitonic Coulomb attraction. Altogether, the two effects induce opposite energy shift at the two valleys, in contrast to the prior red-detuned optical Stark effect that occurs at only one valley \cite{Sie15, Kim14}. This behavior is analogous to the Zeeman effect, which splits \textit{antisymmetrically} the electronic valleys under applied magnetic field \cite{Li14, MacNeill15, Aivazian15, Srivastava15}. We may therefore call this new phenomenon a \textit{Zeeman-type optical Stark effect}, in which the circularly polarized light plays the role of the magnetic field that breaks time-reversal symmetry and lifts the valley degeneracy. This new finding offers much insight into coherent light-matter interactions in TMD materials, and may find important applications in the design of TMD-based photonic and valleytronic devices.

\vspace{0.5cm}
\noindent
\textbf{Supporting Information}
\newline
Transient absorption spectroscopy setup; Optical contributions from the coherent and incoherent effects; Time-trace fitting decomposition analysis; Four-level Jaynes-Cummings model for the optical Stark effect; Fitting analysis based on the Jaynes-Cummings model; Zeeman-type optical Stark effect.

\vspace{0.5cm}
\noindent
\textbf{Acknowledgments}
\newline
The authors acknowledge technical assistance by Qiong Ma and Yaqing Bie during the absorption measurement. N.G. and E.J.S. acknowledge support from the U.S. Department of Energy, BES DMSE (experimental setup and data acquisition) and from the Gordon and Betty Moore Foundation's EPiQS Initiative through Grant GBMF4540 (data analysis and manuscript writing). J.K. acknowledges support from STC Center for Integrated Quantum Materials, NSF Grant No. DMR-1231319 (material growth). Y.-H.L. thanks the funding support from AOARD grant (co-funded with ONRG) FA2386-16-1-4009, Ministry of Science and Technology (MoST 105-2112-M-007-032-MY3; MoST 105-2119-M-007-027) and Academia Sinica Research Program on Nanoscience and Nanotechnology, Taiwan (material growth).


\begin{thebibliography}{}
\bibitem{derivation}  The shifted absorption spectrum has the form of $\alpha(\omega,\Delta E) = \alpha(\omega-\Delta E)$. Hence, the absorption change before and after the  shift is given by $\Delta\alpha(\omega,\Delta E) = \alpha(\omega-\Delta E) - \alpha(\omega) = -\left((\alpha(\omega)-\alpha(\omega-\Delta E))/\Delta E\right)\Delta E$, which can be written as $\Delta\alpha(\omega,\Delta E) = -\left(d\alpha/d\omega\right)\Delta E$.
\end{thebibliography}
\end{document}